\documentclass[epj]{svjour}

\usepackage{amsfonts,amsmath}

\newcommand{\beq}{\begin{equation}}
\newcommand{\eeq}{\end{equation}}

\newcommand{\bsp}{\begin{split}}
	
	\usepackage{epsfig,bbm,cancel,ulem}
	\usepackage[breaklinks=true]{hyperref}
	\usepackage{xcolor}
	\usepackage{wasysym}
	\usepackage[mathscr]{eucal}
	\usepackage{multirow}
	\usepackage{amsmath}
	\usepackage{enumitem}
	\usepackage{appendix}
	\usepackage{slashed}

\begin{document}
		
		\title{Bogomol'nyi equations for Dirac-Born-Infeld cosmic string}
		\author{H.~S.~Ramadhan\inst{1}$^\dagger$, M.~N.~Athaullah\inst{1}$^\flat$ \and I.~Prasetyo\inst{2}$^\sharp$
			\thanks{Corresponding author\\
					$^\dagger$\email{hramad@sci.ui.ac.id}\\
					$^\flat$\email{m.naufal14@ui.ac.id}\\
					$^\sharp$\email{ilham.prasetyo@sampoernauniversity.ac.id}}%
		}                     
		%
		%
		\institute{Departemen Fisika, FMIPA, Universitas Indonesia, Depok 16424, Indonesia \and Department of General Education, Faculty of Art and Sciences, Sampoerna University, Jakarta 12780, Indonesia }
		\date{Received: date / Revised version: date}
		%
		\abstract{
        We revisit the question of whether Dirac–Born–Infeld (DBI) cosmic strings can admit Bogomol’nyi–Prasad–Sommerfield (BPS) configurations. Earlier work by Babichev et al. concluded that DBI strings with the standard Mexican-hat potential possess no BPS limit, implying an unavoidable nonzero binding energy. In contrast, using the BPS Lagrangian method, we show that DBI strings do admit BPS solutions, provided the potential is chosen self-consistently. Imposing the existence of Bogomol’nyi equations uniquely determines the admissible potential and yields exact first-order BPS equations for DBI vortices. We independently verify the consistency of these equations using the stressless (vanishing-pressure) condition on the energy–momentum tensor. The resulting solutions saturate the Bogomol’nyi bound, exhibit zero binding energy, and smoothly recover the Nielsen–Olesen string in the limit $\alpha \to 0$. Regularity of the gauge-field equation requires $\alpha<\pi^2$. A notable outcome of the construction is that the BPS-compatible potential takes a trigonometric form closely related to the sine–Gordon potential, revealing a natural correspondence between the sine–Gordon string and the BPS DBI string. The BPS tension scales linearly with the winding number $n$ but acquires an $\alpha$-dependent deformation.
			%
			\PACS{
				{PACS-key}{discribing text of that key}   \and
				{PACS-key}{discribing text of that key}
			} 
		} 
		\maketitle
\section{Introduction}\label{sec1}

Topological defects, such as cosmic strings, naturally emerge in a variety of high-energy physics and cosmological settings. In particular, brane inflation models predict the formation of cosmic superstrings—including F-strings, D-strings, and their bound states—toward the end of inflationary dynamics~\cite{Cicoli:2023opf,Dasgupta:2002ew,Dasgupta:2004dw}. Similarly, D-term inflation models in supersymmetric and supergravity theories give rise to D-term strings \cite{Binetruy:1996xj,Dvali:2003zh}, which share structural similarities with the standard Abelian–Higgs strings in their low-energy limit. The production of such objects is considered an inevitable outcome of brane collisions \cite{Dvali:1998pa,Sarangi:2002yt,Tye:2004muc}, with their tensions potentially lying near current observational bounds, $G\mu\sim10^{-7}$. This places them within reach of detection in cosmic microwave background anisotropies and stochastic gravitational wave backgrounds \cite{Ma:2008rf,Lin:2022ldu}.

The canonical framework for describing cosmic strings is the Abelian–Higgs model, where the Nielsen–Olesen vortex solution provides the archetypal finite-energy string configuration \cite{Nielsen:1973cs}. The Bogomol’nyi–Prasad–Sommerfield (BPS) limit of this model not only simplifies the equations of motion to first order but also ensures stability, with the string tension scaling linearly with the winding number~\cite{Bogomolny:1975de,Prasad:1975kr}. However, when embedding cosmic strings in brane or string theory contexts, one is often led to nonlinear generalizations of the Abelian–Higgs action. Such extensions may suffer from theoretical issues, including the absence of finite-energy solutions~\cite{Moreno:1998vy,Brihaye:2001ag} or the appearance of multi-valued field configurations \cite{Yang:2000uj}.

Motivated by the effective action on the tachyon kink and vortex~\cite{Sen:2003tm}, Babichev et al. \cite{Babichev:2008qv} employed the Dirac– Born–Infeld (DBI) action, originating from D-brane dynamics, to construct the DBI cosmic string model. The DBI framework provides a natural nonlinear extension of the standard Abelian–Higgs action. The model is free from the aforementioned pathologies and smoothly reduces to the standard Abelian-Higgs cosmic strings in the low-energy limit. The nonlinearity is governed by a single parameter $\alpha$. Yet, unlike the standard Abelian–Higgs string, the DBI cosmic string does not generally admit a BPS limit under the usual Mexican hat potential: the binding energy remains non-vanishing, implying the loss of BPS saturation. This naturally raises two fundamental questions: (i) does a genuine BPS limit exist for the DBI cosmic string, and (ii) if so, what form of scalar potential is required to achieve it? 

In the last two decades, several systematic approaches have been developed to obtain BPS equations for nonlinear field theories. Beyond the classic Bogomol’nyi decomposition \cite{Bogomolny:1975de}, methods such as the strong necessary condition~\cite{Sokalski:2001con}, the first-order formalism~\cite{Bazeia:2005tj,Bazeia:2007df,Bazeia:2017nas}, the on-shell method \cite{Atmaja:2014fha,Atmaja:2015lia}, the the First-Order Euler–Lagrange (FOEL) formalism~\cite{Adam:2016ipc}, and the BPS Lagrangian method~\cite{Atmaja:2015umo} have been successfully applied to vortices~\cite{Atmaja:2015umo,Acalapati:2023rxf,PrasetyaTama:2021elj}, mono-poles and dyons~\cite{Atmaja:2018cod,Atmaja:2020iti,Gunawan:2024pvc,Prasetyo:2018yzl}, or Skyrmions~\cite{Fadhilla:2020rig}.

The purpose of this paper is to derive the BPS equations for the DBI cosmic string using the BPS Lagrangian method. In addition to this primary derivation, we show that the resulting first-order equations are fully consistent with those obtained from the stressless (stress-tensor) condition, thereby providing an independent cross-check of the BPS structure. The paper is organized as follows. In Sec.~\ref{sec2}, we briefly review the standard Abelian–Higgs string. Sec.~\ref{sec3} outlines the BPS Lagrangian method. In Sec.~\ref{sec4}, we apply the formalisms to the DBI cosmic string and present the corresponding BPS equations and tension. Finally, Sec.~\ref{sec6} summarizes our results and discusses their implications.

\section{Nielsen-Olesen Cosmic String}
\label{sec2}

The Nielsen–Olesen vortex line is the string solution of the standard Abelian–Higgs model~\cite{Nielsen:1973cs}, where a complex scalar field $\Phi$ is minimally coupled to an Abelian gauge field $A_\mu$. In $(3+1)$-dimensional Minkowski spacetime with cylindrical coordinates $(t,r,\theta,z)$ and metric signature $(-+++)$, the action takes the form
\begin{equation}\label{action0}
	S=-\int d^4x\sqrt{-g}\left[(D_\mu\Phi)^{\dagger}(D^\mu\Phi)+\frac{1}{4}F_{\mu\nu}F^{\mu\nu}+V(|\Phi|)\right],
\end{equation}
with covariant derivative $D_\mu\equiv\partial_\mu-iqA_\mu$ and gauge field strength tensor $F_{\mu\nu}\equiv\partial_\mu A_\nu-\partial_\nu A_\mu$. The potential is taken to be the familiar symmetry-breaking (Mexican hat) form
\begin{eqnarray}
	V(|\Phi|)=\frac{\lambda}{4}\left(|\Phi|^2-\eta^2\right)^2,
	\label{nomexicanhat}
\end{eqnarray}
where $\lambda$ is the Higgs self-coupling and $\eta$ sets the vacuum expectation value (VEV) of the scalar field. 

To obtain finite-energy string solutions, we impose the cylindrically symmetric ansatz
\begin{equation}
	\Phi=\eta f(\rho)e^{in\theta},\quad a(\rho)=n-qA_\theta(\rho),
	\label{ansatz}
\end{equation}
where $n\in\mathbb{Z}$ is the winding number and $\rho\equiv \sqrt{\lambda}\,\eta r$ is the dimensionless radial coordinate. The appropriate boundary conditions are
\begin{equation}
	\lim_{\rho\to0}f=0, \quad \lim_{\rho\to0}a=n, \quad 
	\lim_{\rho\to\infty}f=1, \quad \lim_{\rho\to\infty}a=0,
	\label{boundary}
\end{equation}
ensuring regularity at the origin and approach to the vacuum at spatial infinity.

With these choices, the energy per unit length (string tension) becomes
\begin{eqnarray}\label{abelian}
	\mu=2\pi\eta^2\int_0^{+\infty}d\rho\, \rho
	&&\left[
	\left(\frac{df}{d\rho}\right)^2
	+\frac{a^2f^2}{\rho^2}
	+\frac{\beta}{\rho^2}\left(\frac{da}{d\rho}\right)^2
	\right.\nonumber\\&&\left.
	+\frac{1}{4}(f^2-1)^2
	\right],
\end{eqnarray}
where $\beta\equiv\lambda/2q^2=m_s^2/m_g^2$ is the dimensionless coupling, with $m_s=\sqrt{\lambda}\eta/\sqrt{2}$ the Higgs mass and $m_g=q\eta$ the vector boson mass. The fields equations follow from the effective Lagrangian density,
\begin{equation}
	\mathcal{L}=-\left(\frac{df}{d\rho}\right)^2
	-\frac{a^2f^2}{\rho^2}
	-\frac{\beta}{\rho^2}\left(\frac{da}{d\rho}\right)^2
	-\frac{1}{4}(f^2-1)^2,
	\label{eq:example1}
\end{equation}
yielding
\begin{eqnarray}
	\frac{d^2f}{d\rho^2}+\frac{1}{\rho}\frac{df}{d\rho}
	-\frac{a^2f}{\rho^2}
	-\frac{1}{2}f(f^2-1)&=&0,\\
	\frac{d^2a}{d\rho^2}-\frac{1}{\rho}\frac{da}{d\rho}
	-\frac{1}{\beta}af^2&=&0.
\end{eqnarray}

At the critical coupling $\beta=1$, the system admits a BPS limit \cite{Bogomolny:1975de,Prasad:1975kr}, where the tension \eqref{abelian} can be rearranged via the Bogomol’nyi decomposition as
\begin{eqnarray}\label{bpstension}
	\mu=&&2\pi\eta^2\int_0^{+\infty}d\rho\, \rho
	\nonumber\\\times&&
	\left\{
	\left(\frac{df}{d\rho}-\frac{af}{\rho}\right)^2
	+\left(\frac{1}{\rho}\frac{da}{d\rho}-\frac{1}{2}(f^2-1)\right)^2
	\right.\nonumber\\&&\left.
	+\frac{1}{\rho}\frac{d}{d\rho}\left[a(f^2-1)\right]
	\right\}.
\end{eqnarray}
Minimizing the energy by setting the quadratic terms to zero leads to the first-order BPS equations,
\begin{eqnarray}
	\frac{df}{d\rho}&=&\frac{af}{\rho},\label{firsteq1}\\
	\frac{da}{d\rho}&=&\frac{\rho}{2}(f^2-1).\label{firsteq2}
\end{eqnarray}
These equations admit smooth vortex solutions with quantized magnetic flux. The corresponding minimal string tension is
\begin{equation}
	{\mu}_\text{BPS}
	=2\pi\eta^2\int_0^{+\infty}d\rho\,\rho
	\left\{
	\frac{1}{\rho}\frac{d}{d\rho}\left[a(f^2-1)\right]
	\right\}
	=2\pi\eta^2n,
	\label{nobpstension}
\end{equation}
demonstrating exact linearity in the winding number $n$.  

The condition $\beta=1$ defines the BPS limit of the Abelian–Higgs string. Physically, this limit corresponds to the case where the Higgs and gauge masses are equal, and the forces between strings cancel out exactly. From the perspective of multi-vortex configurations, the BPS condition guarantees that the binding energy $\mu(n=2k)-2\mu(n=k)=0, \forall\ k\in\mathbb{Z}$, ensuring that vortices neither attract nor repel~\cite{Taubes:1979tm,Jaffe:1980mj}.

\section{BPS Lagrangian Method}
\label{sec3}

In this section, we revisit the BPS Lagrangian method in its original construction~\cite{Atmaja:2015umo}. This method provides a systematic way to obtain Bogomol'nyi equations by introducing a modified Lagrangian density, called the BPS Lagrangian $\mathcal{L}_{\text{BPS}}$, such that
\begin{equation}\label{bpslagrangianeq}
	\mathcal{L} - \mathcal{L}_{\text{BPS}} = 0 .
\end{equation}
The BPS Lagrangian is defined in terms of the BPS energy function 
\begin{equation}
Q=-\int d^Dx\,\mathcal{L}_\text{BPS}
\end{equation}
, with the BPS energy (or, in the string context, the BPS tension $\mu_{\text{BPS}}$) expressed as the boundary difference of $Q$:
\begin{equation}
	\mu_{\text{BPS}} = \lim_{r\to\infty} Q - \lim_{r\to 0} Q = \int_{r\to 0}^{r\to\infty} dQ .
\end{equation}

In many cases, the BPS energy function can be expressed as separable functions of effective fields $\tilde{\phi}i$,
\begin{equation}
Q = \prod_{i=1}^N \varPhi_i(\tilde{\phi}_i).
\end{equation}
For a cylindrically symmetric and static configuration, the string tension is given by the total energy,
\begin{equation}
\mu = -k \int d\rho\ \rho\ \mathcal{L},
\end{equation}
where $k$ is a normalization constant. Motivated by this structure, we assume a separable ansatz for the BPS energy function,
\begin{equation}
\label{Q}
Q = k\ F(f)\ A(a),
\end{equation}
which implies that the BPS Lagrangian satisfies
\begin{equation}
Q=-k\int d\rho \ \rho\ \mathcal{L_\text{BPS}}.
\end{equation}
This leads to a BPS Lagrangian of the form
\begin{equation}
\label{bpslagrangian}
\mathcal{L}_{\text{BPS}} = -\frac{Q_f}{\rho} f'(\rho) - \frac{Q_a}{\rho} a'(\rho),
\end{equation}
where $Q_f \equiv F'(f)A(a)$ and $Q_a\equiv F(f)A'(a)$. The condition~\eqref{bpslagrangianeq} enforces quadratic consistency conditions, which can be solved to determine both the BPS equations and the corresponding scalar potential $V(f)$.

To illustrate this method, consider the effective Lagrangian of the Abelian--Higgs system,
\begin{equation}
	\mathcal{L} = -f'(\rho)^2 - \frac{a^2 f^2}{\rho^2} - \frac{\beta}{\rho^2} a'(\rho)^2 - V(f),
\end{equation}
with $V(f)$ now is a dimensionless and undetermined scalar potential. 
It can easily be seen  that in this case, $Q=2\pi\eta^2FA$. 
Inserting $\mathcal{L}_{\text{BPS}}$ into $\mathcal{L} - \mathcal{L}_{\text{BPS}}$ gives a quadratic equation in $f'$ or $a'$,
\begin{equation}
	-f'(\rho)^2-\frac{a^2f^2}{\rho^2}-\frac{\beta}{\rho^2}a'(\rho)^2-V + \frac{Q_f}{\rho} f'(\rho) + \frac{Q_a}{\rho} a'(\rho)=0.
\end{equation} 
Solving for $f'$,
\begin{equation}
	f'=\frac{1}{2\rho}\left(Q_f\pm\sqrt{Q_f^2-4a^2f^2-4\rho^2V-4\beta a'^2+4\rho Q_aa'}\right).\label{f'nolbps}
\end{equation}
Demanding uniqueness of solutions (so that $f'$ has a single root) leads to
\begin{equation}
	a'=\frac{1}{2\beta}\left\{\rho Q_a\pm\sqrt{\rho^2Q_a^2+ \beta\left(Q_f^2-4 a^2f^2-4\rho^2V\right)}\right\}.\label{a'nolbps}
\end{equation}
The uniqueness of the root of $a'$ implies algebraic constraints $\beta  \left(Q_f^2-4 a^2 f^2\right)+\rho ^2 \left(Q_a^2-4 \beta  V\right)=0$, which yields
\begin{eqnarray}
	Q_a^2-4\beta V&=&0,\ \ \ \
	Q_f^2-4 a^2 f^2=0.
\end{eqnarray}
Since
\begin{eqnarray}
	Q_f=F'(f) A(a)&=&\pm2af,
\end{eqnarray}
we may choose
\begin{eqnarray}
	A(a)=c_aa,\qquad F(f)=\pm \frac{1}{c_a}\left(f^2+c_f\right),
\end{eqnarray}
where $c_a$ and $c_f$ are constants.
These yield
\begin{eqnarray}
Q_a&=&F(f)A'(a)=\pm\left(f^2+c_f\right),\\
    V(f)&=&\frac{1}{4\beta}Q_a^2=\frac{1}{4\beta}\left(f^2+c_f\right)^2.
\end{eqnarray}
We can set $c_f=-1$ to satisfy the requirement that the potential needs to have minima at $|f|=1$. Following Eqs. \eqref{f'nolbps}-\eqref{a'nolbps}, the BPS equations becomes
\begin{eqnarray}
	f' &=& \pm \frac{af}{\rho},\ \ \ \ 
	a' = \pm \frac{\rho}{2\beta} (f^2 - 1).
\end{eqnarray}
The corresponding energy bound is then
\begin{eqnarray}
	\mu_\text{BPS}=&& {2\pi \eta^2} FA\bigg|^{\rho\to\infty}_{\rho=0} =\pm {2\pi \eta^2} a(f^2-1)\bigg|^{\rho\to\infty}_{\rho=0} 
	\nonumber\\=&& \pm {2\pi \eta^2}n ,
\end{eqnarray}
saturating exactly the BPS limit when taking the positive sign.
Note that if we set $\beta = 1$, the equations reduce to the ordinary Nielsen-Olesen string,
\begin{eqnarray}
	a'&=&\frac{\rho}{2}(f^2-1),\quad\quad\quad
	f'=\frac{af}{\rho},\nonumber\\
	V&=&\frac{1}{4}(f^2-1)^2,\quad
	\mu_\text{BPS}=2\pi\eta^2n.
\end{eqnarray}
\section{DBI Cosmic String}
\label{sec4}

Babichev et al.~\cite{Babichev:2008qv} considered the DBI-inspired action
\begin{eqnarray}
	S=-T\int d^4x && \left\{\sqrt{-\det{\left[g_{\mu \nu}+(\mathcal{D}_{(\mu}\Phi)(\mathcal{D}_{\nu)}\Phi)^{\dagger}+\ell_s^2\mathcal{F}_{\mu\nu}\right]}}
		\right.\nonumber\\&&\left.
		-\sqrt{-g}+\sqrt{-g}{V(|\Phi|)}\right\},\label{dbiaction}
\end{eqnarray}
where $T$ is the brane tension, $\ell_s$ denotes the fundamental string length, $g\equiv\det\left(g_{\mu\nu}\right)$, and $V(|\Phi|)$ is a dimensionless scalar potential. The covariant derivative and gauge field strength are, repsectively,
\begin{equation}
	\mathcal{D}_\mu \equiv \partial_\mu - i\hat{q}\mathcal{A}_\mu, 
	\qquad 
	\mathcal{F}_{\mu\nu} \equiv \partial_\mu \mathcal{A}_\nu - \partial_\nu \mathcal{A}_\mu.
\end{equation}

After rescaling to dimensionless variables,
\begin{eqnarray}
	&&\hat{\Phi}\equiv\frac{\Phi}{\ell_s},~~
	\hat{\mathcal{F}_{\mu\nu}}\equiv\ell_s^2\mathcal{F}_{\mu\nu},~~
	\hat{\mathcal{D}_{\mu}}\equiv\ell_s\mathcal{D}_{\mu},~~
	\nonumber\\&&
	\hat{r}\equiv\frac{r}{\ell_s},~~
	\hat{\eta}\equiv\frac{\eta}{\sqrt{T}\ell_s},~~\hat{\lambda}\equiv\lambda T\ell_s^4,\label{rescale}
\end{eqnarray}
the action becomes
\begin{eqnarray}
	S=-T\ell_s^4\int d^4\hat{x}&&\Bigg\{\sqrt{-\det{\left[
				\hat{g}_{\mu\nu}
	+(\hat{\mathcal{D}}_{(\mu}\hat{\Phi})(\hat{\mathcal{D}}_{\nu)}\hat{\Phi})^{\dagger}+\hat{\mathcal{F}}_{\mu\nu}\right]}}
	\nonumber\\&&
	-\sqrt{-
		\hat{g}
	}+\sqrt{-
		\hat{g}
	}\ V(|\hat{\Phi}|)\Bigg\},
    \label{eq:SDBI}
\end{eqnarray}
with $\hat{g}_{\mu\nu}$ denoting the dimensionless metric tensor.
Note that here we do not impose any specific form of the scalar potential $V(|\hat{\Phi}|)$. Instead, its structure will be determined self-consistently through the BPS Lagrangian method, as required by the existence of Bogomol’nyi equations.

Assuming a static, cylindrically symmetric ansatz (with further rescaling  $\rho\equiv\hat{\lambda}^{1/2}\hat{\eta}\hat{r}$)
\begin{equation}
	\hat{\Phi}=\hat{\eta}f(\rho)e^{in\theta},~~~~~~~~~~~~
	a(\rho)=n-\hat{q}\hat{\mathcal{A}}_\theta(\rho),\label{dbiansatz}
\end{equation}
with boundary condition
\noindent\begin{eqnarray}
	\lim_{\rho\to0}f=0,~~~
	\lim_{\rho\to0}a=n,~~~
	\lim_{\rho\to\infty}f=1,~~~
	\lim_{\rho\to\infty}a=0,
\end{eqnarray}
tension $\mu=-S/\ell_s^2d\hat{t}d\hat{z}$ for this string become
\noindent\begin{eqnarray}
	\mu=&&\frac{4\pi\eta^2}{\alpha}\int_0^{+\infty} d\rho\,\rho
	\nonumber\\\times&&
	\Bigg\{\sqrt{\left(1+\alpha f'(\rho)^2\right)\left(1+\alpha\frac{a^2f^2}{\rho^2}\right)+\frac{\alpha\beta}{\rho^2}a'(\rho)^2}
	\nonumber\\&&
	-1+V(f)\Bigg\},\label{dbitension}
\end{eqnarray}
with $\beta=\hat{\lambda}/2\hat{q}^2=\lambda/2q^2$ is coupling constant as in Nielsen-Olesen, dan $\alpha\equiv2\hat{\lambda}\hat{\eta}^4$ is parameter of deformation from standard Abelian-Higgs string, which at limit $\alpha\to0$, the string reduced into standard Abelian-Higgs string.

With effective Lagrangian
\begin{eqnarray}
	\mathcal{L}=&&-\sqrt{\left(1+\alpha f'(\rho)^2\right)\left(1+\alpha\frac{a^2f^2}{\rho^2}\right)+\frac{\alpha\beta}{\rho^2}a'(\rho)^2}
	\nonumber\\&&
	+1-V(f),\label{dbilagrangian}
\end{eqnarray}
the corresponding Euler-Lagrange field equations become
\begin{eqnarray}
	\frac{d}{d\rho}\left[\rho\gamma f'(\rho)\left(1+\alpha\frac{a^2f^2}{\rho^2}\right)\right]&&
	\nonumber\\
	-\gamma\frac{a^2f}{\rho}\left(1+\alpha f'(\rho)^2\right)-\frac{\rho}{\alpha}V'(f)&&=0,\label{fel}\\
	\frac{d}{d\rho}\left(\frac{\gamma}{\rho}a'(\rho)\right)-\frac{\gamma}{\beta}\frac{af^2}{\rho}\left(1+\alpha f'(\rho)^2\right)&&=0,\label{ael}
\end{eqnarray}
with
\begin{eqnarray}
	\gamma\equiv\frac{1}{\displaystyle{\sqrt{\left(1+\alpha f'(\rho)^2\right)\left(1+\alpha\frac{a^2f^2}{\rho^2}\right)+\frac{\alpha\beta}{\rho^2}a'(\rho)^2}}}.
\end{eqnarray}
The presence of the square-root structure in \eqref{dbilagrangian}, inherited from the DBI action, significantly modifies the standard Abelian–Higgs equations of motion~\cite{Nielsen:1973cs}. In particular, the $\gamma$ factor plays a role analogous to a Lorentz-type suppression factor, encoding the nonlinear kinetic corrections that become important for large gradients of the fields~\cite{Brown:2007zzh}. In the next two sub-sections we shall apply two BPS methods to obtaining the Bogomolnyi equations for DBI cosmic string.

\subsection{Bogomol’nyi Construction via BPS Lagrangian Method}

Since the string tension is related to the effective Lagrangian $\mathcal{L}$ through
\begin{eqnarray}
	\mu=\frac{4\pi\eta ^2}{\alpha}\int_0^{+\infty}d\rho\ \rho\ \mathcal{L},
\end{eqnarray}
we adopt a BPS energy function of the separable form
\begin{eqnarray}
	\label{QBPS}
	Q=\frac{4\pi\eta ^2}{\alpha}F(f)A(a).
\end{eqnarray}

Following the BPS Lagrangian method, Eq.~\eqref{bpslagrangianeq} leads to the constraint
\begin{eqnarray}
	-\sqrt{\left(1+\alpha f'^2\right)\left(1+\alpha\frac{a^2f^2}{\rho^2}\right)+\frac{\alpha\beta}{\rho^2}a'^2}+1-V
	\nonumber\\
	+\frac{Q_f}{\rho}f'+\frac{Q_a}{\rho}a'=0,
\end{eqnarray}
which can be viewed as a quadratic equation for $f'$. Its general solution takes the form
\begin{eqnarray}
	f'=\frac{-\displaystyle{\frac{2Q_f}{\rho}}\left(1-V+\displaystyle{\frac{Q_a}{\rho}}a'\right)\pm\sqrt{D}}{2\left[\displaystyle{\frac{Q_f^2}{\rho^2}}-\alpha\left(1+\alpha\displaystyle{\frac{a^2f^2}{\rho^2}}\right)\right]},\label{f'}
\end{eqnarray}
with discriminant
\begin{eqnarray}
	D&\equiv&-4\left[\displaystyle{\frac{Q_f^2}{\rho^2}}-\alpha\left(1+\alpha\displaystyle{\frac{a^2f^2}{\rho^2}}\right)\right]\left[\left(1-V+\frac{Q_a}{\rho}a'\right)^2
	\right.\nonumber\\&&\left.
	-1-\alpha\frac{a^2f^2}{\rho^2}-\frac{\alpha\beta}{\rho^2}a'^2\right]\nonumber\\
	&&+\left[2\displaystyle{\frac{Q_f}{\rho}}\left(1-V+\displaystyle{\frac{Q_a}{\rho}}a'\right)\right]^2.\label{Ddbi}
\end{eqnarray}
To ensure the BPS structure, we require $D=0$, which guarantees that $f'$ admits a unique solution. Substituting this condition back into Eq.~\eqref{Ddbi} yields the corresponding expression for $a'$,
\begin{equation}
	a'=\frac{-\left(1+\alpha\dfrac{a^2f^2}{\rho^2}\right)\dfrac{Q_a}{\rho}(1-V)\pm\sqrt{DD}}{\beta\dfrac{Q_f^2}{\rho^4}+\left(1+\alpha\dfrac{a^2f^2}{\rho^2}\right)\left(\dfrac{Q_a^2-\alpha\beta}{\rho^2}\right)},\label{a'}
\end{equation}
with
\begin{eqnarray}
	DD&&\equiv-\left(1+\alpha\displaystyle{\frac{a^2f^2}{\rho^2}}\right)\Biggl\{\nonumber\beta\left[\displaystyle{\frac{Q_f^2}{\rho^4}}-\frac{\alpha}{\rho^2}\left(1+\alpha\displaystyle{\frac{a^2f^2}{\rho^2}}\right)\right]
	\nonumber\\	&&\times
	\left[(1-V)^2-1-\alpha\frac{a^2f^2}{\rho^2}+\frac{1}{\alpha}\frac{Q_f^2}{\rho^2}\right]\nonumber\\
	&&+\left(1+\alpha\displaystyle{\frac{a^2f^2}{\rho^2}}\right)\displaystyle{\frac{Q_a^2}{\rho^2}}\left[-1-\alpha\frac{a^2f^2}{\rho^2}+\frac{1}{\alpha}\frac{Q_f^2}{\rho^2}\right]\Biggr\}.
\end{eqnarray}
Analogous to the condition imposed on $f'$, we must also require $DD=0$ in order to obtain a unique solution for $a'$. This condition leads to the constraint 
\begin{eqnarray}
	0=&&-\frac{1}{\rho^2}\left\{\alpha\beta\left[(1-V)^2-1\right]+Q_a^2\right\}
	\nonumber\\&&
	-\frac{1}{\rho^4}\Bigg\{\beta\left(Q_f^2-\alpha^2a^2f^2\right)+Q_a^2\alpha a^2f^2\nonumber\\
	&&-\frac{1}{\alpha}\left\{\alpha\beta\left[(1-V)^2-1\right]+Q_a^2\right\}\left(Q_f^2-\alpha^2a^2f^2\right)\Bigg\}\nonumber\\
	&&+\frac{1}{\alpha\rho^6}\left(Q_f^2-\alpha^2a^2f^2\right)\left[\beta\left(Q_f^2-\alpha^2a^2f^2\right)
	\right.\nonumber\\&&\left.
	+Q_a^2\alpha a^2f^2\right].\label{DDdbi}
\end{eqnarray}
For Eq.~\eqref{DDdbi} to hold for all values of $\rho$, the coefficients of each power of $\rho$ must vanish independently. This requirement yields two key constraints
\begin{eqnarray}
	\alpha\beta\left[\left(1-V(f)\right)^2-1\right]+Q_a^2&=&0,\label{order-2dbi}\\
	\beta\left(Q_f^2-\alpha^2a^2f^2\right)+Q_a^2\alpha a^2f^2&=&0.\label{order-4dbi}
\end{eqnarray}
Eq.~\eqref{order-2dbi} implies that $Q_a=F(f)A'(a)$ must be independent of $a$. Consequently, we may choose
\begin{eqnarray}
	A'(a)=c_1,\qquad A(a)=c_1a+c_0,
\end{eqnarray}
where $c_1$ and $c_0$ are constants.
These yield
\begin{eqnarray}
	Q_a=F(f)c_1,\quad Q_f=F'(f)\left(c_1a+c_0\right).\label{qfqa}
\end{eqnarray}
Substituting Eq.~\eqref{qfqa} into Eq.~\eqref{order-4dbi} yields
\begin{eqnarray}
	F'=\frac{dF}{df}=\pm\frac{\sqrt{\alpha\beta-F^2c_1^2}}{c_1a+c_0}\sqrt{\frac{\alpha}{\beta}}af,
\end{eqnarray}
which can be integrated to obtain
\begin{eqnarray}
	F=s_F\ \frac{\sqrt{\alpha\beta}}{c_1}
	\sin\left[\sqrt{\frac{\alpha}{\beta}}\left(\frac{1}{2}f^2+c_2\right)\right],\label{Fdbi}
\end{eqnarray}
where $s_F\equiv\pm 1$, $c_2$ is an integration constant, and $c_0=0$ for consistency. Thus,
\begin{eqnarray}
	Q_a=&&s_F\ \sqrt{\alpha\beta}
	\sin\left[\sqrt{\frac{\alpha}{\beta}}\left(\frac{1}{2}f^2+c_2\right)\right],\ \ \ 
	\label{qa2}\\
	Q_f=&&s_F\ \alpha a f
	\cos\left[\sqrt{\frac{\alpha}{\beta}}\left(\frac{1}{2}f^2+c_2\right)\right].\label{qf2}
\end{eqnarray}


We can now determine the shape of the potential. From Eqs.~\eqref{order-2dbi}, \eqref{qa2}, and \eqref{qf2}, the potential is given by
\begin{eqnarray}
	V=1
	+s_V\cos\left[\sqrt{\frac{\alpha}{\beta}}\left(\frac{1}{2}f^2+c_2\right)\right]
	\label{V},
\end{eqnarray}
with $s_V\equiv\pm1$.
Substituting Eqs.~\eqref{qa2}, \eqref{qf2}, and \eqref{V} into Eq.~\eqref{a'} yields
\begin{eqnarray}
	a'=-s_F\ s_V\ \frac{\rho\left(1+\alpha\displaystyle{\frac{a^2f^2}{\rho^2}}\right)}{\sqrt{\alpha\beta}}\tan\left[\sqrt{\frac{\alpha}{\beta}}\left(\frac{1}{2}f^2+c_2\right)\right].
	\nonumber\\
	\label{eqa'}
\end{eqnarray}
Substituting Eqs.~\eqref{qa2}, \eqref{qf2},  \eqref{V}, and \eqref{eqa'} into Eq.~\eqref{f'} further gives
\begin{eqnarray}
	f'=-s_Fs_V\frac{af}{\rho}.\label{eqf'}
\end{eqnarray}
The asymptotic behavior of $a$ requires
\begin{eqnarray}
	\lim_{\rho\to\infty}a'=-s_Fs_V\frac{\rho}{\sqrt{\alpha\beta}}\tan\left[\sqrt{\frac{\alpha}{\beta}}\left(\frac{1}{2}+c_2\right)\right]=0,\label{limita'}
\end{eqnarray} 
which is satisfied for
\begin{eqnarray}
	c_2=-\frac{1}{2}+\sqrt{\frac{\beta}{\alpha}}m\pi, \qquad m\in\mathbb{Z}.
\end{eqnarray}
Substituting this relation back into Eqs.~\eqref{V} and \eqref{eqa'}, we obtain
\begin{eqnarray}
	  V&=&1+(-1)^m\ s_V\ \cos\left(\frac{1}{2}\sqrt{\frac{\alpha}{\beta}}(f^2-1)\right),\nonumber\\
a'&
=&
-s_F\ s_V\frac{\rho\left(1+\alpha\displaystyle{\frac{a^2f^2}{\rho^2}}\right)}{\sqrt{\alpha\beta}}\tan\left(\frac{1}{2}\sqrt{\frac{\alpha}{\beta}}\left(f^2-1\right)\right).\nonumber\\
\end{eqnarray}
To ensure regularity of the gauge-field profile, we must avoid the divergence of the tangent function appearing in $a'$. Since $f^2-1$ varies within the interval $-1 \le f^2-1 \le 0$, this requires  $\sqrt{\alpha/\beta}\left(f^2-1\right)/2>-\pi/2$. This leads to the restriction
\begin{eqnarray}
	  0\leq\frac{\alpha}{\beta}<\pi^2\simeq9.8696.
\end{eqnarray}
For $V$ to represent a symmetry-breaking potential, it must possess minima at $|f|=1$, with no additional extrema in the interval $-1<f<1$ other than a local maximum at $f=0$. 
From the first derivative,
\begin{eqnarray}    
    V'(f)=-(-1)^ms_V\sqrt{\frac{\alpha}{\beta}}f\sin\left(\frac{1}{2}\sqrt{\frac{\alpha}{\beta}}\left(f^2-1\right)\right),
\end{eqnarray}
it can be inferred that $\alpha/\beta\leq 4\pi^2$. Evaluating the second derivative
\begin{eqnarray}    
    V''(f)&=&-(-1)^ms_V\Bigg[\sqrt{\frac{\alpha}{\beta}}\sin\left(\frac{1}{2}\sqrt{\frac{\alpha}{\beta}}\left(f^2-1\right)\right)\nonumber\\
    &&+\frac{\alpha}{\beta}f^2\cos\left(\frac{1}{2}\sqrt{\frac{\alpha}{\beta}}\left(f^2-1\right)\right)\Bigg],
\end{eqnarray}
to ensure the extrema at $|f|=1$ are minima, we must choose $(-1)^ms_V=-1$. Alongside the restriction of $\alpha/\beta$, these also ensure the extremum at $f=0$ is local maximum.



From the BPS energy function in Eq.~\eqref{QBPS},
\begin{eqnarray}
	Q&&=\frac{4\pi\eta^2}{\alpha}FA
	\nonumber\\
	&&=(-1)^ms_F\ 4\pi\eta^2a\sqrt{\frac{\beta}{\alpha}}\sin\left(\frac{1}{2}\sqrt{\frac{\alpha}{\beta}}(f^2-1)\right),
\end{eqnarray}
the BPS string tension follows as
\begin{equation}
	\mu_\text{BPS}=4\pi\eta^2n\sqrt{\frac{\beta}{\alpha}}\sin\left(\frac{1}{2}\sqrt{\frac{\alpha}{\beta}}\right),\label{tensiondbibps0}
\end{equation}
where $(-1)^m s_F=1$ to ensure the positivity of $\mu_\text{BPS}$. The BPS equations and the potential simplify to
\begin{eqnarray}
	\label{a'0}
	a'&=&\frac{\rho\left(1+\alpha\displaystyle{\frac{a^2f^2}{\rho^2}}\right)}{\sqrt{\alpha\beta}}\ \tan\left(\frac{1}{2}\sqrt{\frac{\alpha}{\beta}}(f^2-1)\right),\nonumber\\
	f'&=&\frac{af}{\rho},\nonumber\\
	V&=&1-\cos\left(\frac{1}{2}\sqrt{\frac{\alpha}{\beta}}(f^2-1)\right).
    \label{eq:BPSlagr0}
\end{eqnarray}

\subsection{Bogomol’nyi Construction via Stressless Method}

In addition to the BPS Lagrangian method discussed previously, the Bogomol’nyi equations can also be derived by imposing the stressless condition on the energy–momentum tensor. This approach, originally applied to Abelian–Higgs vortices in Ref.~\cite{deVega:1976xbp}, exploits the fact that for static, cylindrically symmetric configurations the conservation law $\nabla_{\mu}T^{\mu\nu}=0$ reduces to
\begin{equation}
\label{constraint}
\partial_1(x^1 T^{11})=(x^1)^2 T^{22},    \end{equation}
which implies the first-order constraints
\begin{equation}
\label{stressless}
T^{11}=T^{22}=0,   
\end{equation}
Using the definition of the energy–momentum tensor, $T^{\mu\nu}=2/\sqrt{-g}\ \delta S/\delta \hat{g}_{\mu\nu}$, one obtains~\cite{Cordero:2007rw}
\begin{eqnarray}    
    T^{\mu\nu}&=&-T\ell_s^4 \left[\frac{\sqrt{-M}}{\sqrt{-g}}(M^{-1})^{(\mu\nu)}-(1-V)\hat{g}^{\mu\nu}\right],\nonumber\\
    M_{\mu\nu}&=& \hat{g}_{\mu\nu}+(\hat{\mathcal{D}}_{(\mu}\Phi)(\hat{\mathcal{D}}_{\nu)}\Phi)^\dagger+ \hat{\mathcal{F}}_{\mu\nu},\nonumber\\
(M^{-1})^{(\mu\nu)}&=&\frac{(M^{-1})^{\mu\nu}+(M^{-1})^{\nu\mu}}{2},
\end{eqnarray}
where $M\equiv\det\left(M_{\mu\nu}\right)$. Evaluating the components for the standard vortex ansatz yields
\begin{eqnarray}    
    T^{00}&=&-T\ell_s^4  \left(-\frac{1}{\gamma}+1-V\right),\nonumber\\
    T^{\rho\rho}&=&-T\ell_s^4 \frac{\alpha}{2\hat{\eta}^2}\left[\gamma\left(1+\alpha \frac{a^2 f^2}{\rho^2}\right) +V-1\right],\nonumber\\
    T^{\theta\theta}&=&-T\ell_s^4 \frac{\alpha}{2\hat{\eta}^2}\frac{1}{\rho^2}\left[\gamma\left(1+\alpha f'(\rho)^2\right)+V-1\right],\nonumber\\
    T^{zz}&=&-T^{00}.
\end{eqnarray}

Applying the constraints \eqref{stressless}, one can combine the components through $T^{\rho\rho}+\rho^2T^{\theta\theta}=0$ and $T^{\rho\rho}-\rho T^{\theta\theta}=0$, which leads to the first-order system 
\begin{eqnarray}    
    f'&=&s_f\frac{af}{\rho},\label{eq:solVS1}\\
    a'&=&s_a\frac{\rho}{\sqrt{\alpha\beta}}\frac{\sqrt{1-(1-V)^2}}{1-V}\left(1+\alpha \frac{a^2 f^2}{\rho^2}\right),
    \label{eq:solVS2}
\end{eqnarray} 
where $s_f\equiv\pm1$ and $s_a\equiv \pm1$. Note that the condition \eqref{stressless} enforces $V-1<0$.

Substituting Eqs.~\eqref{eq:solVS1}–\eqref{eq:solVS2} back into the Euler–Lagrange equation \eqref{ael} gives
\begin{eqnarray}    
    -\frac{af}{\rho}\frac{(1-V)}{\sqrt{\alpha\beta}}\left[\displaystyle{\sqrt{\frac{\alpha}{\beta}}}f-\displaystyle{\frac{s_fs_a}{\sqrt{1-(1-V)^2}}\frac{dV}{df}}\right]=0,
\end{eqnarray}
which yields a first-order differential equation for $V(f)$. Integrating, we obtain
\begin{eqnarray}    
    V(f)=1-\cos\left[\sqrt{\frac{\alpha}{\beta}}\left(\frac{1}{2}f^2+c\right)\right],
\end{eqnarray}
where $c$ is an integration constant. Since the potential must have minima at $|f|=1$ and local maximum at $f=0$, it is easy to show that $c$ satisfies 
\begin{eqnarray}
	c=-\frac{1}{2}+\sqrt{\frac{\beta}{\alpha}}m\pi, \qquad m\in\mathbb{Z}.
\end{eqnarray}
The potential therefore becomes
\begin{eqnarray}    
    V=1-\cos\left(\frac{1}{2}\sqrt{\frac{\alpha}{\beta}}\left(f^2-1\right)\right).\label{V2}
\end{eqnarray}
Finally, the stressless condition requires $V-1<0$, hence
\begin{eqnarray}    
    V-1=-\cos\left(\frac{1}{2}\sqrt{\frac{\alpha}{\beta}}\left(f^2-1\right)\right)<0.
\end{eqnarray}
This implies the bound  
\begin{equation}
\alpha/\beta<\pi^2.
\end{equation}
This bound guarantees that the potential has no additional extrema within $-1\leq f\leq1$, so that $f=0$ remains local maximum and $|f|=1$ is the only minima.


Substituting the potential \eqref{V2} into Eq.~\eqref{eq:solVS2} gives
\begin{eqnarray}
    a'&=&s_a\frac{\rho\left(1+\alpha \displaystyle{\frac{a^2 f^2}{\rho^2}}\right)}{\sqrt{\alpha\beta}}\frac{\left|\displaystyle{\sin\left(\frac{1}{2}\sqrt{\frac{\alpha}{\beta}}\left(f^2-1\right)\right)}\right|}{\displaystyle{\cos\left(\frac{1}{2}\sqrt{\frac{\alpha}{\beta}}\left(f^2-1\right)\right)}}\nonumber\\
    &=&-s_a\frac{\rho\left(1+\alpha \displaystyle{\frac{a^2 f^2}{\rho^2}}\right)}{\sqrt{\alpha\beta}}\tan\left(\frac{1}{2}\sqrt{\frac{\alpha}{\beta}}\left(f^2-1\right)\right),\label{bpsa}
\end{eqnarray}
where $\sin\left[\sqrt{\alpha/\beta}\left(f^2-1\right)/2\right]<1$ in the interval $-1\leq f\leq1$ for $\alpha/\beta<\pi^2$. This same bound also guarantees that the denominator $\cos\left[\sqrt{\alpha/\beta}\left(f^2-1\right)/2\right]$ never vanishes within $-1\leq f\leq1$, avoiding singularity
at $\alpha/\beta=\pi^2$.

The tension can now be expressed as
\begin{eqnarray}
	\mu=&&\frac{4\pi\eta^2}{\alpha}\int_0^{+\infty} d\rho\,\rho\,\Bigg\{\Bigg[\alpha\left(f'-s_f\frac{af}{\rho}\right)^2
	\nonumber\\
    &&+\Bigg(\left(1+s_f\alpha f'\frac{af}{\rho}\right)\sin{F}+s_a\frac{\sqrt{\alpha\beta}}{\rho}a'\cos{F}\Bigg)^2\nonumber\\
	&&+\Bigg(\left(1+s_f\alpha f'\frac{af}{\rho}\right)\cos{F}-s_a\frac{\sqrt{\alpha\beta}}{\rho}a'\sin{F}\Bigg)^2\Bigg]^{\frac{1}{2}}\nonumber\\
    &&-\cos{F}\Bigg\},\label{completesquaretension}
\end{eqnarray}
with
\begin{eqnarray}
	F\equiv&&\frac{1}{2}\sqrt{\frac{\alpha}{\beta}}\left(f^2-1\right).
\end{eqnarray}
Substituting the first-order relations \eqref{eq:solVS1} and \eqref{bpsa} into Eq.~\eqref{completesquaretension} yields the BPS tension in the form
\begin{eqnarray}
	\mu_\text{BPS}=&&\frac{4\pi\eta^2}{\alpha}\int_0^{+\infty} d\rho\,\rho\
	\nonumber\\
    &&\Bigg\{s_f\frac{\sqrt{\alpha\beta}}{\rho}\frac{d}{d\rho}\left[a\sin\left(\frac{1}{2}\sqrt{\frac{\alpha}{\beta}}\left(f^2-1\right)\right)\right]\nonumber\\
    &&-(s_f+s_a)\frac{\sqrt{\alpha\beta}}{\rho}a'\sin\left(\frac{1}{2}\sqrt{\frac{\alpha}{\beta}}\left(f^2-1\right)\right)\Bigg\}.\nonumber\\
\end{eqnarray}
With some algebraic manipulations and integration techniques, one finds
\begin{eqnarray}
	\mu_\text{BPS}=&&\frac{s_f-s_a}{1-\displaystyle{\frac{s_f}{s_a}}}4\pi\eta^2\sqrt{\frac{\beta}{\alpha}}a\sin\left(\frac{1}{2}\sqrt{\frac{\alpha}{\beta}}\left(f^2-1\right)\right)\Bigg|^{\rho\to\infty}_{\rho\to0}\
	\nonumber\\
    =&&\frac{s_f-s_a}{1-\displaystyle{\frac{s_f}{s_a}}}4\pi\eta^2n\sqrt{\frac{\beta}{\alpha}}\sin\left(\frac{1}{2}\sqrt{\frac{\alpha}{\beta}}\right).
\end{eqnarray}
Requiring that the BPS tension be finite and positive uniquely selects $s_f=1$ and $s_a=-1$. With this sign choice, the BPS equations and the corresponding tension reduce to
\begin{eqnarray}
a'&=&\frac{\rho\left(1+\alpha\displaystyle{\frac{a^2f^2}{\rho^2}}\right)}{\sqrt{\alpha\beta}}\ \tan\left(\frac{1}{2}\sqrt{\frac{\alpha}{\beta}}(f^2-1)\right),\nonumber\\
	f'&=&\frac{af}{\rho},\nonumber\\
\mu_\text{BPS}&=&4\pi\eta^2n\sqrt{\frac{\beta}{\alpha}}\sin\left(\frac{1}{2}\sqrt{\frac{\alpha}{\beta}}\right).
\end{eqnarray}


\subsection{BPS Limit and DBI Deformation}

To understand how the DBI deformation modifies the vortex structure, let us first examine the limit $\alpha\to0$. In this regime, the DBI action reduces to the standard Abelian–Higgs model and the BPS equations smoothly approach
\begin{eqnarray}
	\lim_{\alpha\to0}a'&=&\frac{\rho}{2\beta}(f^2-1),\quad\quad\quad
	\lim_{\alpha\to0}f'=\frac{af}{\rho},\nonumber\\
	\lim_{\alpha\to0}V&=&\frac{1}{8}\frac{\alpha}{\beta}(f^2-1)^2,\quad
	\lim_{\alpha\to0}\mu_\text{BPS}=2\pi\eta^2n.
\end{eqnarray}
To ensure that the potential reduces to the usual Mexican-hat form, independent of $\beta$, we impose $\beta=1$,
\begin{eqnarray}
a'&=&\frac{\rho\left(1+\alpha\displaystyle{\frac{a^2f^2}{\rho^2}}\right)}{\sqrt{\alpha}}\tan\left(\frac{\sqrt{\alpha}}{2}(f^2-1)\right),\label{a'3}\\
	f'&=&\frac{af}{\rho},\label{f'3}\\
	V&=&1-\cos\left(\frac{\sqrt{\alpha}}{2}(f^2-1)\right),\label{dbipotentialbps}\\
	\mu_\text{BPS}&=&\frac{4\pi\eta^2n}{\sqrt{\alpha}}\sin\left(\frac{\sqrt{\alpha}}{2}\right).\label{dbitensionbps}
\end{eqnarray}

The fact that the potential is independent of $\beta$ implies that the entire DBI system reduces continuously to the Nielsen-Olesen string for any allowed value of $\beta$. 
The parameter $\alpha$ remains bounded by 
$\alpha<\pi^2$
, beyond which the string solution ceases to exist. Note that Eq.~\eqref{f'} is identical to the BPS equation for $f'$ in the Nielsen–Olesen string~\cite{Nielsen:1973cs}, and coincides with the Born– Infeld–Higgs string studied in \cite{Shiraishi:1990zi}.

\begin{figure*}[h]\sidecaption
	\centering
	\includegraphics[width=0.8\textwidth]{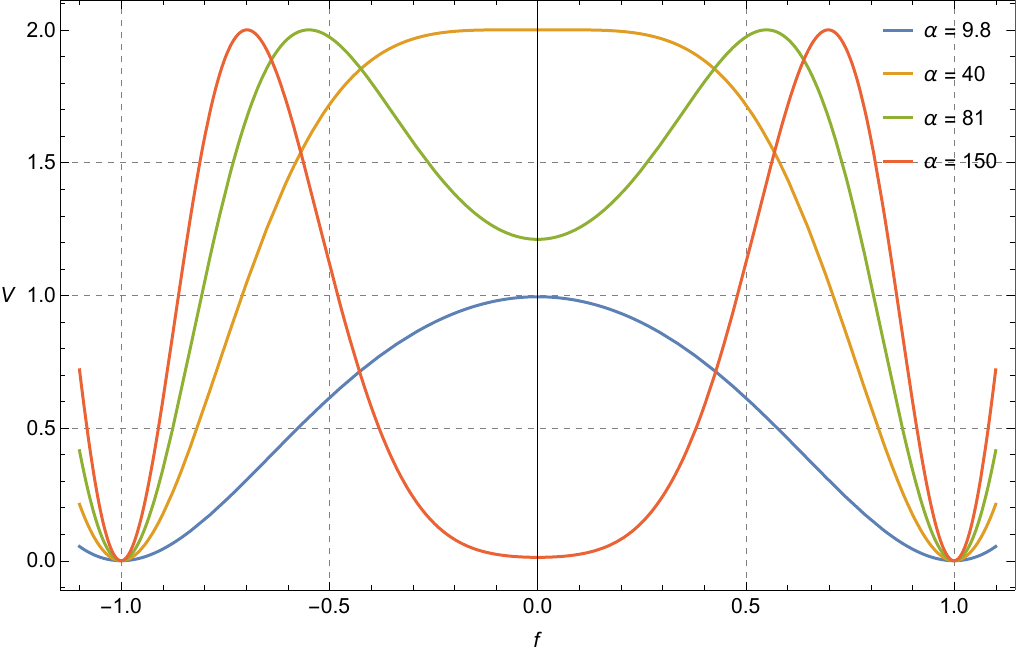}
	\caption{Plot of the potential in Eq.~\eqref{dbipotentialbps} for for different values of $\alpha$. For $\alpha < 4\pi^2$, the potential maintains the desired symmetry-breaking structure with minima at $|f|=1$. As $\alpha$ increases beyond the critical value $4\pi^2 \simeq 39.478$, local minimum appears in the interval $-1<f<1$, destabilizing the topological sector.}
	\label{fig:potential}
\end{figure*}

A striking feature of our construction is that the resulting potential takes a trigonometric form. Expanding in small $\alpha$, one recovers the standard Mexican-hat potential as the leading-order approximation,
\begin{equation}
	V\approx\frac{\alpha}{8}\left(f^2-1\right)^2+\mathcal{O}(\alpha^2)=\frac{\hat{\lambda}}{4}(|\hat{\Phi}|^2-\hat{\eta}^2)^2+\mathcal{O}\left(\alpha^2\right).
\end{equation}
At the same time, the exact structure closely resembles the well-known sine–Gordon (SG) potential, 
\begin{equation}
	V_{\text{SG}}(\phi)=1-\cos(\phi),
\end{equation}
under the field redefinition $\phi \sim\sqrt{\alpha}(f^2-1)/2$. The SG model is an integrable 1+1 dimensional field theory that supports topological solitons and breather excitations~\cite{Rajaraman:1982is}. Its potential, $U(\phi) \equiv 1 - \cos(\phi)$, supports topological solitons (kink and antikink) solutions, which interpolate between adjacent minima at $\phi = 2\pi n$ and $\phi = 2\pi (n+1)$, where $n \in \mathbb{Z}$. Similar cosine-type structures are also known to emerge in DBI tachyon condensation, where kink configurations follow sine–Gordon–like dynamics~\cite{Sen:2002nu,Canfora:2005mt}. Some  In Fig.~\ref{fig:potential}, we plot $V(f)$ for representative values of $\alpha$. 
When $\alpha$ exceeds $4\pi^2$, the cosine deformation becomes strong enough to generate an extra local minimum within the interval $-1<f<1$. This additional vacuum spoils the standard symmetry-breaking structure, destabilizing the topological sector, and preventing the existence of stable vortex solutions. Interestingly, this instability arises at a value much larger than the upper bound $\alpha=\pi^2$, indicating that the singular behavior of the gauge-field BPS equation originates from the structure of the Bogomol'nyi completion rather than from the potential itself. The stressless method, in contrast, constrains the potential through energy–momentum conservation and naturally reproduces the identical regularity bound as that implied by the gauge-field equation.

\begin{figure*}[h]\sidecaption
	\centering
	\includegraphics[width=0.8\textwidth]
	{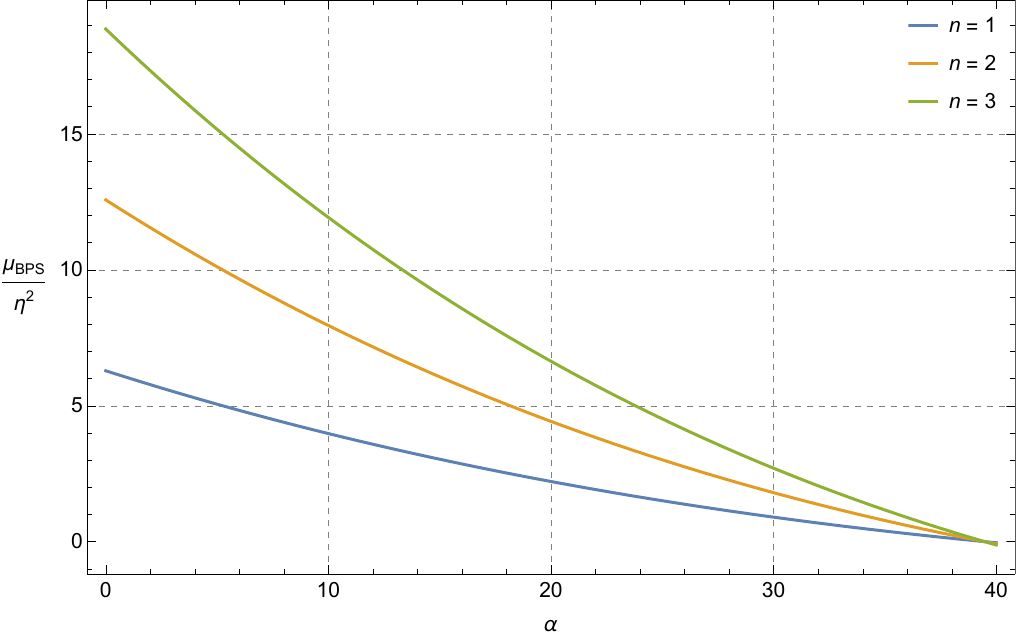}
	\caption{Plot of the BPS tension from Eq.~\eqref{dbitensionbps} for various winding number $n$. The tension decreases with increasing $\alpha$ and vanishes at the critical value $\alpha = 4\pi^2 \simeq 39.478$. Beyond this point, the tension formally becomes negative
    .}
	\label{fig:mubps}
\end{figure*}

While the BPS tension continues to scale linearly with the winding number $n$, it acquires a trigonometric dependence on $\alpha$, introducing a nontrivial deformation to the usual Abelian–Higgs scaling. As a result, depending on the value of $\alpha$, the DBI string tension can be either enhanced or suppressed relative to the standard case. 
This behavior is reminiscent of cosmic F- and D-strings~\cite{Copeland:2003bj,Polchinski:2004ia}, where the effective four-dimensional tension depends sensitively on compactification details and warp factors (e.g., $\mu_{4D} = e^{2A}\mu_{10D}$), allowing strong suppression of the effective tension relative to the fundamental one. Our DBI BPS result displays an analogous nontrivial rescaling, but with a distinct dynamical origin: here, the trigonometric dependence emerges directly from the nonlinear structure of the DBI kinetic term, rather than from moduli or warping effects in higher-dimensional compactifications. In Fig.~\ref{fig:mubps}, we plot the dependence of $\mu_\text{BPS}$ on $\alpha$. 
As $\alpha \to \alpha_c\equiv4\pi^2$, the tension approach zero. This critical value coincides with the point at which the potential would develop an unwanted extra local minimum. However, because BPS solutions only exist for $\alpha\geq\pi^2$, the tension in the range for $\pi^2\leq\alpha\leq4\pi^2$ is considered unphysical.

\begin{figure*}[h]\sidecaption
\centering
\includegraphics[width=0.8\textwidth]
{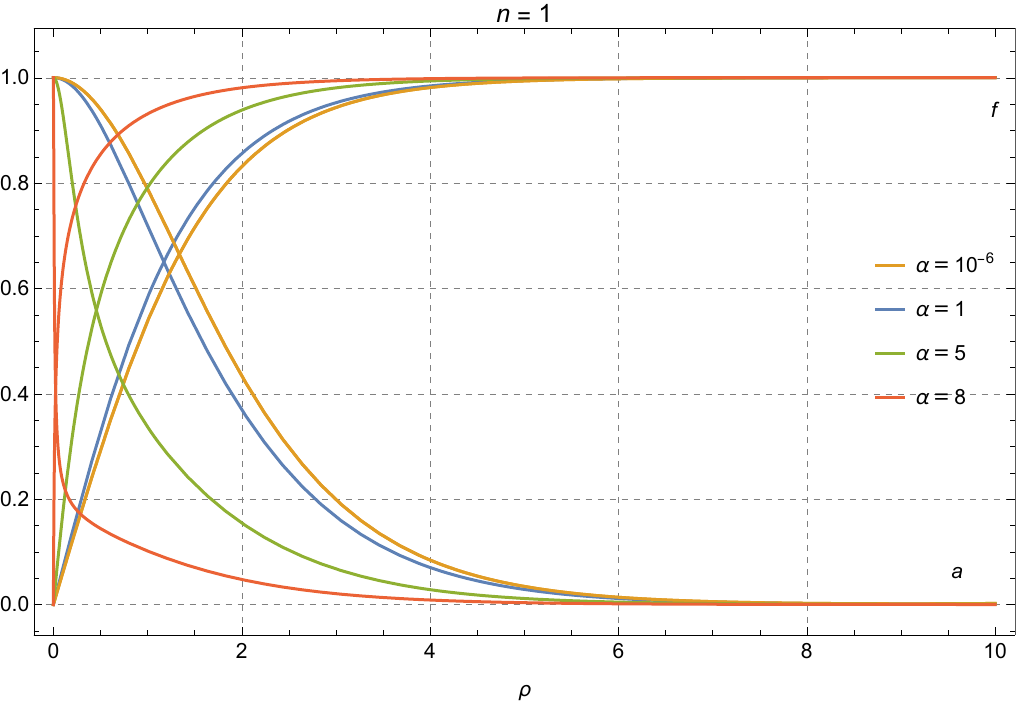}
\caption{Numerical solutions of Eqs.~\eqref{a'}–\eqref{f'} for winding number $n=1$ at various values of $\alpha$. As $\alpha$ increases, the scalar and gauge field profiles become more localized around the string core. At the upper bound 
$\alpha = \pi^2$
, the solutions collapse, and no regular vortex configuration exists beyond this bound.}
\label{fig:vsalphaDBI}
\end{figure*}

As in the case with other BPS vortices, despite being first-order differential equations the analytical solutions are impossible to obtain. In Fig.~\ref{fig:vsalphaDBI}, we present the numerical solutions of the BPS equations for several values of the deformation parameter $\alpha$. As $\alpha$ increases, the scalar and gauge field profiles become increasingly localized around the string core. Due to the theoretical restriction 
$\alpha < \pi^2$
, the solutions exhibit an extreme localization as $\alpha$ approaches this upper bound, beyond which the string configuration ceases to exist. 
Conversely, for small values of $\alpha$, the profiles broaden and become more diffuse, smoothly converging to the familiar Nielsen–Olesen vortex~\cite{Nielsen:1973cs} in the Abelian–Higgs limit as $\alpha \to 0$.

\begin{figure*}[h]\sidecaption
\centering
\includegraphics[width=0.8\textwidth]
{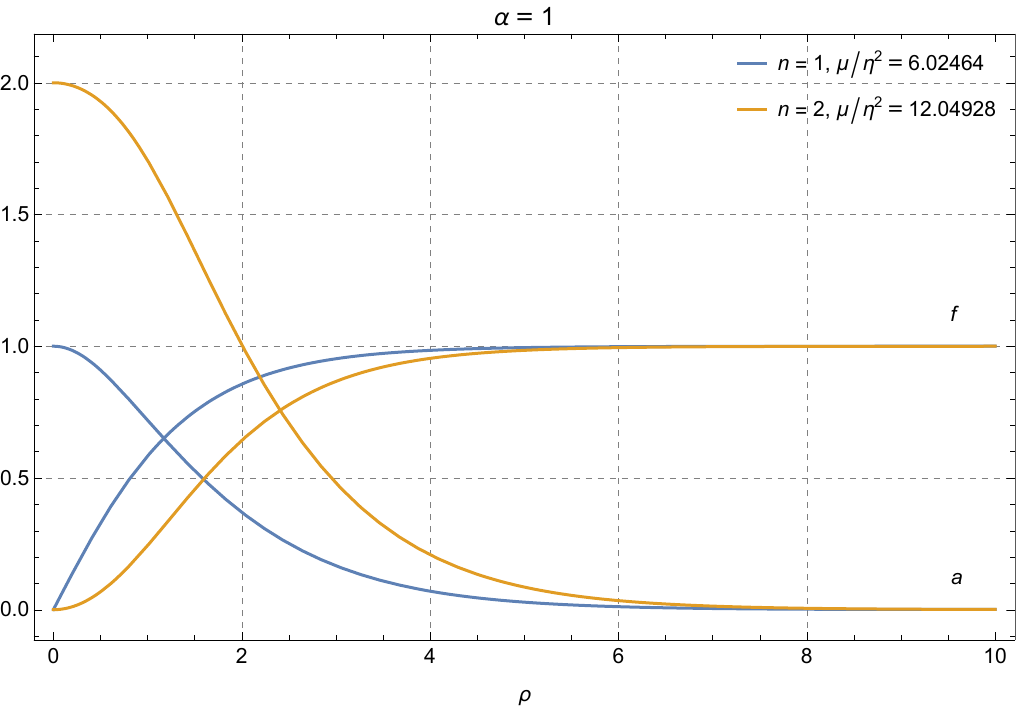}
\caption{Numerical solutions of Eqs.~\eqref{a'}–\eqref{f'} for winding numbers $n=1$ and $n=2$ at $\alpha=1$. The numerically computed tensions are in excellent agreement with the analytic BPS expression \eqref{dbitensionbps}, confirming the linear scaling with $n$ and the vanishing of the binding energy at the BPS state.}
\label{fig:vsnDBI}
\end{figure*}

In Fig.~\ref{fig:vsnDBI}, we show the numerical solutions for winding numbers $n=1$ and $n=2$. The numerical evaluation of the string tension, obtained from direct integration of Eq.~\eqref{dbitension}, agrees with the analytical BPS tension formula \eqref{dbitensionbps} to at least five decimal places, providing a strong consistency check of the BPS Lagrangian construction. Furthermore, the binding energy, computed as $\mu(n=2)-2\mu(n=1)$, vanishes, demonstrating that the multi-winding solutions saturate the BPS bound. This result is consistent with analyses of BPS Abelian-Higgs vortices~\cite{Taubes:1979tm,deVega:1976xbp}, where the string tension scales linearly with the winding number, reflecting the absence of binding energy between multi-winding configurations.

Our result challenges the earlier claim of Babichev et al.~\cite{Babichev:2008qv}, who argued that DBI cosmic strings cannot possess a BPS limit and instead necessarily exhibit a positive binding energy. Their conclusion was derived under the assumption of the Mexican-hat potential, within which the nonlinear square-root structure of the DBI action indeed obstructs a Bogomol’nyi completion of the energy functional. In contrast, by employing the BPS Lagrangian method we have shown that the potential needs not be fixed a priori but can be determined consistently from the requirement of Bogomol’nyi equations. This procedure naturally yields a class of symmetry-breaking potentials that admit genuine BPS DBI vortex solutions, with vanishing binding energy and an analytically determined BPS tension, Eq.~\eqref{dbitensionbps}. The construction reduces smoothly to the standard Abelian–Higgs BPS limit as $\alpha\to0$. Thus, contrary to the conclusions of Babichev et al., the existence of BPS DBI cosmic strings is not precluded in general, but rather depends crucially on the potential chosen; an insight made transparent by the BPS Lagrangian framework.




\section{Conclusions}\label{sec6}

Babichev et.~al.~\cite{Babichev:2008qv} studied a class of high-energy cosmic string inspired by $D$-brane dynamics and originally argued that DBI cosmic strings cannot saturate the Bogomol’nyi bound because the square-root structure of the DBI action obstructs the standard Bogomol’nyi completion under the usual Mexican-hat potential. Their analysis suggested that DBI strings necessarily exhibit nonzero binding energy and thus deviate from the BPS paradigm of Abelian–Higgs vortices.

In this work, we demonstrated that this claim is not generic: by allowing the scalar potential to be determined self-consistently from the requirement of Bogomol’nyi equations, we derived explicit BPS equations for DBI vortices. We employ the BPS Lagrangian method~\cite{Atmaja:2015umo}, and the resulting configurations saturate the BPS bound, possess vanishing binding energy, and admit an exact analytic expression for the tension, Eq.~\eqref{dbitensionbps}, which smoothly reduces to the Nielsen–Olesen result in the limit $\alpha \to 0$.

To achieve our objective, we employ two methods: the BPS Lagrangian and the stressless condition methods. Assuming the BPS energy density is separable, the BPS Lagrangian method gives transparent route to the first-order equations and the analytic expression for the BPS tension, requiring only systematic identification of the total-derivative structure. This method is particularly convenient for constructing BPS solutions in noncanonical or generalized models. The stressless method, on the other hand, relies on imposing vanishing pressure conditions on the energy-momentum tensor, ensuring the configuration satisfies energy-momentum conservation. While this requirement is quite general, it involves more cumbersome calculations, including decomposing the energy density into complete-square terms, which can be more technically demanding.

A distinctive feature of our results is the emergence of a trigonometric potential, Eq.~\eqref{dbipotentialbps}, which closely resembles the sine–Gordon (SG) potential. This connection suggests that DBI cosmic strings can be viewed as higher-dimensional analogues of SG solitons, potentially sharing stability and spectral properties with them. 
DBI/tachyon effective actions are known to produce trigonometric or periodic effective potentials for kink-type excitations in reduced descriptions, and field-theoretic models of tachyon kinks have been shown to give rise to sine-Gordon–like dynamics under appropriate approximations~\cite{Canfora:2005mt}. In cosmology, some papers study SG-type potentials for axion fields, where kinks can be stretched into string-like structures~\cite{Yoshino:2012kn}.

Another significant result is the exact BPS tension, Eq.~\eqref{dbitensionbps},
which reveals a trigonometric deformation of the usual linear scaling. It reduces to the Abelian–Higgs result $\mu_{\text{BPS}}=2\pi\eta^2 n$ in the $\alpha\to0$ limit. Depending on $\alpha$, the tension can be either enhanced or suppressed relative to the Abelian–Higgs case. It formally vanishes at $\alpha=4\pi^2$, however it ceases to be physical at $\alpha=\pi^2$ due to the regularity condition of the BPS solutions
. This dynamical suppression is reminiscent of tension rescaling in cosmic F- and D-strings~\cite{Copeland:2003bj,Polchinski:2004ia}, though with a distinct origin rooted in the DBI nonlinearity rather than in warping or compactification effects. In comparison, the required potential in the Born-Infeld-Higgs model \cite{Shiraishi:1990zi} also have some restrictions on its nonlinear parameter, as the potential have complex value when the restrictions are not satisfied.

In brane inflation scenarios, cosmic strings are expected to inherit DBI-type dynamics from their D-brane origin. The existence of BPS DBI strings with well-defined Bogomolnyi equations and analytic tension thus provides atheoretical foundation for studying their stability, interactions, and possible observational signatures. The trigonometric deformation of the BPS tension suggests that deviations from the Abelian–Higgs spectrum could leave imprints in gravitational wave backgrounds~\cite{Damour:2004kw,Kibble:2004hq,Hindmarsh:2011qj}.

\section*{Acknowledgement}
We thank Leonardus Putra for the fruitful discussions on D-brane dynamics. IP is funded by Sampoerna University's Internal Research Grant No.~003/IRG/SU/AY. 2023-2024.

\begin{appendix}

\end{appendix}


\end{document}